\documentstyle[twoside,fleqn,espcrc2]{article}
\newcommand{\Real}{\mbox{I \hspace{-0.82em} R}}
\newcommand{\Complex}{
       \mbox{C \hspace{-1.16em} \raisebox{-0.018em}{\sf l}}\;}
\newcommand{\eto}[2]{\mbox{\large e}^{\raisebox{#1pt}{$#2$}}}
\title{
  \quad\vskip-3.5cm \hfill {\normalsize
  \begin{tabular}[t]{l}
                     \rule{0ex}{1ex}Freiburg THEP-92/26 \\[.0ex]
                     \rule{0ex}{1ex}Tel Aviv TAUP 2007-92 \\[.0ex]
                     \rule{0ex}{1ex}November 1992
  \end{tabular}}
  \vskip1.2cm  
Monte Carlo Simulations of 2-Dimensional Quantum Gravity
Coupled to $c = 1$ Matter
   \thanks  {To appear in {\em Lattice '92}, Amsterdam 1992,
             eds.\ P. van Baal and J. Smit,
             Nucl.\ Phys.\ B (Proc.\ Suppl.).}
}
\author{Thomas Filk\address{University of Freiburg,
                    Department of Physics, Hermann-Herder-Str. 3,
                    D-7800 Freiburg, Germany},
Mihail Marcu\address{School of Physics and Astronomy,
                     Raymond and Beverly Sackler Faculty of Exact Sciences,
                     Tel Aviv University, 69978 Tel Aviv, Israel}
and Bernhard Scheffold$^{\rm a}$}

\begin{document}

\begin{abstract}
We present results of a high precision Monte Carlo simulation of
dynamically triangulated random surfaces (up to $\approx$ 34,000
triangles) coupled to one scalar field ($c=1$).
The mean square extent has been
measured for different actions to test the universality of the
leading term as a function of the size of the surfaces. Furthermore,
the integrated 2-point correlation function for vertex operators
is compared with conformal field theory and matrix model predictions.
\end{abstract}

\maketitle

\section{INTRODUCTION}

This talk reports about the results of several projects, where -- apart
from the authors -- also M. Agishtein, R. Ben-Av, I. Klebanov,
A. A. Migdal and S. Solomon were involved. Additional details
and even more precise data will be published soon.

$c=1$ matter coupled to 2-dimensional quantum gravity seems to be a
critical model in the sense that most approaches to the theory of
matter coupled to gravity in 2 dimensions predict qualitatively
different behaviour for $c<1$ and $c>1$. Furthermore, there are two
different theoretical approaches (conformal field theory
\cite{Polyakov,KPZ,David,DK,Polchinski} and matrix models
\cite{GM,DS,BK,Gross,Kostov,GKN})
which give quite detailed predictions for this model.

The main aim of our high precision Monte Carlo simulations was
to confirm those predictions where conformal field theory and matrix
models agree, and to decide or clarify the situation where the two
approaches differ.

\section{DEFINITION OF THE MODEL AND THEORETICAL PREDICTIONS}

Let ${\cal T}_N$ denote the set of abstract triangulations of the
2-dimensional
sphere with $N$ triangles. This is equivalent to the set of planar,
regular graphs of degree 3 with $N$ vertices and without non-trivial
2-point subgraphs. Each graph can be characterized by its adjacency
matrix: $C_{ij}=1$ if $i$ is a neighbour of $j$ and $C_{ij}=0$
otherwise. The partition function of the models we simulated is
\begin{eqnarray}
  Z_N &=& \sum_{{\cal T}_N} \int \! dX_1 \ldots dX_N ~ \\[-0.3cm]
  & & \hspace{0.7cm}
     \exp \left(-\frac{1}{2} \sum_{i,j} C_{ij} E(X_i,X_j) \right)~~.
     \nonumber
\end{eqnarray}
$X_i$ is a real, scalar field attached to the faces of the
triangulation. We mainly considered
\begin{equation}
\label{action2}
         E_2(X,Y) ~=~ \frac{1}{2} (X-Y)^2 ~~,
\end{equation}
the discrete analog of the squared derivative, but for a check of
universal properties of some of the quantities we also present data for
\begin{equation}
\label{action1}
        E_1(X,Y) ~=~ |X-Y| ~~.
\end{equation}
Both actions are expected to describe $c=1$ matter coupled to
quantum gravity.

The 2-point function for the tachyon with momentum $p$ is
\begin{eqnarray}
         G_N(p) &=& \left\langle \frac{1}{N^2} \sum_{i,j}
             {\rm e}^{ip(X_i - X_j)} \right\rangle  \\
        &=&  \left\langle \left| \frac{1}{N} \sum_{i}
             {\rm e}^{ipX_i} \right|^2
             \right\rangle ~~.      \nonumber
\end{eqnarray}
{}From this one obtains the moments of $X$ by derivatives, e.g. the
mean square extent
\begin{eqnarray}
    \langle X^2 \rangle_N &=& \frac{1}{N^2} \left\langle
         \sum_{i,j}^{N} (X_i - X_j )^2 \right\rangle \\
            &=&    \label{twopoint}
                - \left.
                \frac{\partial^2}{\partial p^2} G_N (p)
                \right|_{p=0}  ~~.
\end{eqnarray}
For these quantities exist theoretical predictions from conformal
field theory
\cite{Polyakov,KPZ,David,DK,Polchinski} and from matrix models
\cite{GM,DS,BK,Gross,Kostov,GKN}),
which can be summarized in the following formulas:
\vspace{0.2cm}

1.~~Both approaches agree that
\begin{equation}
\label{scale}
   G(p) ~=~ \frac{z}{{\rm sinh}^2 \sqrt{z}}
       \hspace{1.0cm}  z=\frac{3}{2} \langle X^2 \rangle p^2 ~~.
\end{equation}
\vspace{0.2cm}

2.~~Furthermore, both approaches predict the general dependence of
the mean square extent on the number of triangles:
\begin{equation}
\label{mean}
       \langle X^2 \rangle_N ~ \propto  ~ (\ln N)^2 ~~.
\end{equation}
This equation, which has first been conjectured in \cite{km},
implies that $G_N(p)$ becomes a function of $(p \ln N)$ only.
\vspace{0.2cm}

3.~~In order to fix the coefficient in (\ref{mean}) one has to
normalize the field $X$. This is usually done by
using the asymptotic behaviour of the 2-point funcion in flat space:
\begin{eqnarray}
  \langle X_{\sigma} X_{\sigma'} \rangle
  & \longrightarrow & - \alpha' \: \ln|\sigma-\sigma'| \\
    & & {\rm ~~~~~~for~~}
    |\sigma-\sigma'| \rightarrow \infty~~. \nonumber
\end{eqnarray}
For the Gaussian action (\ref{action2}) this normalization is
known:
\begin{equation}
   \alpha'(E_2) ~=~ 1/2\pi ~ \approx ~ 0.159  ~~.
\end{equation}

To determine this factor for action $E_1$ of eq.\ (\ref{action1})
we performed a Monte Carlo simulation on a flat 2-dimensional lattice
(a $512\times 512$ square lattice with periodic boundary conditions) and found
\begin{equation}
    \alpha'(E_1) ~\approx~  0.134(2)  ~~.
\end{equation}
The conformal weight of the tachyon operator in flat space is
\begin{equation}
      \Delta_0(p) ~=~ \frac{\alpha'}{4} p^2 ~~.
\end{equation}
Using the KPZ formula \cite{KPZ} to relate the weight of a conformal
field in flat space with the one coupled to 2-dimensional gravity,
\begin{equation}
    \Delta(p) ~=~
        \frac{\sqrt{1-c+24 \Delta_0(p)}-\sqrt{1-c}}{\sqrt{25-c}
        -\sqrt{1-c}} ~~,
\end{equation}
one can apply formal scaling
arguments to obtain $G(p)$ as a function of the
area $A$ \cite{KPZ,David,DK,Distler}:
\begin{equation}
\label{green}
    G_A(p) ~\propto ~ A^{1 - 2 \Delta(p)}
\end{equation}
For $c=1$ this expression is non-analytic at $p=0$,
and we get a wrong minus sign if we try to
obtain the mean square extent directly
from this formula, according to eq.\ref{twopoint}.
For $p$ large however, equation
(\ref{scale}), which has been derived in conformal field theory
by introducing a cut-off \cite{Polchinski}, agrees with (\ref{green}),
provided that the mean square extent is given by
\begin{equation}
       \langle X^2 \rangle ~=~ \frac{\alpha'}{6} (\ln N)^2 ~~.
\end{equation}
On the other hand, the mean square extent has been calculated in
matrix models \cite{Kostov,GKN}, yielding
\begin{equation}
\label{matrix}
       \langle X^2 \rangle ~=~ \frac{\alpha}{6} (\ln N)^2 ~~.
\end{equation}
This calculation uses the action
\begin{equation}
       E_{\alpha}(X,Y) ~=~ \frac{|X-Y|}{\sqrt{\alpha}}~~.
\end{equation}
Matrix models, however, do not only generate graphs corresponding to
triangulations but also graphs with nontrivial 2-point subgraphs.
This might imply that the ``mass'' $1/\sqrt{\alpha}$ has to be
renormalized before one can compare the results.

\section{DATA AND RESULTS}

We made simulations for $N$ =140, 420, 1260, 3780, 11340 and 34020
triangles. The update of the scalar field was speeded up considerably
by the use of the VMR (``valleys-to-mountains reflections) cluster
algorithm \cite{VMR}. For the update of the triangulation we used
the triangle flip \cite{KKM,BKKM,David2,Billoire,ADF}, which had to be
supplemented by a local update procedure for the scalar field (we used
both heat bath and Metropolis). Careful analysis of the autocorrelations
revealed an autocorrelation time of $\approx 300$ local sweeps (with a
few clusters after each sweep) for the largest system, for which we
performed about 200.000 sweeps.

No errorbars in the figures indicates that the errors
are small compared to the symbols used to mark the data points.
We should emphasize that the data for the simulations of action
$E_1$ are less accurate since we have not finished our runs.

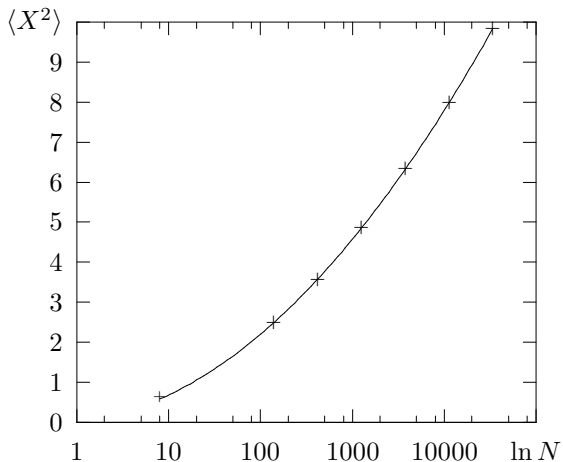
\begin{figure}[htb]
\setlength{\unitlength}{0.240900pt}
\ifx\plotpoint\undefined\newsavebox{\plotpoint}\fi
\sbox{\plotpoint}{\rule[-0.175pt]{0.350pt}{0.350pt}}%
\special{em:linewidth 0.3pt}%
\begin{picture}(1049,900)(200,-100)
\tenrm
\put(264,158){\special{em:moveto}}
\put(985,158){\special{em:lineto}}
\put(264,158){\special{em:moveto}}
\put(284,158){\special{em:lineto}}
\put(985,158){\special{em:moveto}}
\put(965,158){\special{em:lineto}}
\put(242,158){\makebox(0,0)[r]{0}}
\put(264,221){\special{em:moveto}}
\put(284,221){\special{em:lineto}}
\put(985,221){\special{em:moveto}}
\put(965,221){\special{em:lineto}}
\put(242,221){\makebox(0,0)[r]{1}}
\put(264,284){\special{em:moveto}}
\put(284,284){\special{em:lineto}}
\put(985,284){\special{em:moveto}}
\put(965,284){\special{em:lineto}}
\put(242,284){\makebox(0,0)[r]{2}}
\put(264,347){\special{em:moveto}}
\put(284,347){\special{em:lineto}}
\put(985,347){\special{em:moveto}}
\put(965,347){\special{em:lineto}}
\put(242,347){\makebox(0,0)[r]{3}}
\put(264,410){\special{em:moveto}}
\put(284,410){\special{em:lineto}}
\put(985,410){\special{em:moveto}}
\put(965,410){\special{em:lineto}}
\put(242,410){\makebox(0,0)[r]{4}}
\put(264,473){\special{em:moveto}}
\put(284,473){\special{em:lineto}}
\put(985,473){\special{em:moveto}}
\put(965,473){\special{em:lineto}}
\put(242,473){\makebox(0,0)[r]{5}}
\put(264,535){\special{em:moveto}}
\put(284,535){\special{em:lineto}}
\put(985,535){\special{em:moveto}}
\put(965,535){\special{em:lineto}}
\put(242,535){\makebox(0,0)[r]{6}}
\put(264,598){\special{em:moveto}}
\put(284,598){\special{em:lineto}}
\put(985,598){\special{em:moveto}}
\put(965,598){\special{em:lineto}}
\put(242,598){\makebox(0,0)[r]{7}}
\put(264,661){\special{em:moveto}}
\put(284,661){\special{em:lineto}}
\put(985,661){\special{em:moveto}}
\put(965,661){\special{em:lineto}}
\put(242,661){\makebox(0,0)[r]{8}}
\put(264,724){\special{em:moveto}}
\put(284,724){\special{em:lineto}}
\put(985,724){\special{em:moveto}}
\put(965,724){\special{em:lineto}}
\put(242,724){\makebox(0,0)[r]{9}}
\put(264,787){\special{em:moveto}}
\put(284,787){\special{em:lineto}}
\put(985,787){\special{em:moveto}}
\put(965,787){\special{em:lineto}}
\put(242,787){\makebox(0,0)[r]{$\langle X^2 \rangle$}}
\put(264,158){\special{em:moveto}}
\put(264,178){\special{em:lineto}}
\put(264,787){\special{em:moveto}}
\put(264,767){\special{em:lineto}}
\put(264,113){\makebox(0,0){1}}
\put(307,158){\special{em:moveto}}
\put(307,168){\special{em:lineto}}
\put(307,787){\special{em:moveto}}
\put(307,777){\special{em:lineto}}
\put(365,158){\special{em:moveto}}
\put(365,168){\special{em:lineto}}
\put(365,787){\special{em:moveto}}
\put(365,777){\special{em:lineto}}
\put(394,158){\special{em:moveto}}
\put(394,168){\special{em:lineto}}
\put(394,787){\special{em:moveto}}
\put(394,777){\special{em:lineto}}
\put(408,158){\special{em:moveto}}
\put(408,178){\special{em:lineto}}
\put(408,787){\special{em:moveto}}
\put(408,767){\special{em:lineto}}
\put(408,113){\makebox(0,0){10}}
\put(452,158){\special{em:moveto}}
\put(452,168){\special{em:lineto}}
\put(452,787){\special{em:moveto}}
\put(452,777){\special{em:lineto}}
\put(509,158){\special{em:moveto}}
\put(509,168){\special{em:lineto}}
\put(509,787){\special{em:moveto}}
\put(509,777){\special{em:lineto}}
\put(538,158){\special{em:moveto}}
\put(538,168){\special{em:lineto}}
\put(538,787){\special{em:moveto}}
\put(538,777){\special{em:lineto}}
\put(552,158){\special{em:moveto}}
\put(552,178){\special{em:lineto}}
\put(552,787){\special{em:moveto}}
\put(552,767){\special{em:lineto}}
\put(552,113){\makebox(0,0){100}}
\put(596,158){\special{em:moveto}}
\put(596,168){\special{em:lineto}}
\put(596,787){\special{em:moveto}}
\put(596,777){\special{em:lineto}}
\put(653,158){\special{em:moveto}}
\put(653,168){\special{em:lineto}}
\put(653,787){\special{em:moveto}}
\put(653,777){\special{em:lineto}}
\put(683,158){\special{em:moveto}}
\put(683,168){\special{em:lineto}}
\put(683,787){\special{em:moveto}}
\put(683,777){\special{em:lineto}}
\put(697,158){\special{em:moveto}}
\put(697,178){\special{em:lineto}}
\put(697,787){\special{em:moveto}}
\put(697,767){\special{em:lineto}}
\put(697,113){\makebox(0,0){1000}}
\put(740,158){\special{em:moveto}}
\put(740,168){\special{em:lineto}}
\put(740,787){\special{em:moveto}}
\put(740,777){\special{em:lineto}}
\put(797,158){\special{em:moveto}}
\put(797,168){\special{em:lineto}}
\put(797,787){\special{em:moveto}}
\put(797,777){\special{em:lineto}}
\put(827,158){\special{em:moveto}}
\put(827,168){\special{em:lineto}}
\put(827,787){\special{em:moveto}}
\put(827,777){\special{em:lineto}}
\put(841,158){\special{em:moveto}}
\put(841,178){\special{em:lineto}}
\put(841,787){\special{em:moveto}}
\put(841,767){\special{em:lineto}}
\put(841,113){\makebox(0,0){10000}}
\put(884,158){\special{em:moveto}}
\put(884,168){\special{em:lineto}}
\put(884,787){\special{em:moveto}}
\put(884,777){\special{em:lineto}}
\put(942,158){\special{em:moveto}}
\put(942,168){\special{em:lineto}}
\put(942,787){\special{em:moveto}}
\put(942,777){\special{em:lineto}}
\put(971,158){\special{em:moveto}}
\put(971,168){\special{em:lineto}}
\put(971,787){\special{em:moveto}}
\put(971,777){\special{em:lineto}}
\put(985,158){\special{em:moveto}}
\put(985,178){\special{em:lineto}}
\put(985,787){\special{em:moveto}}
\put(985,767){\special{em:lineto}}
\put(985,113){\makebox(0,0){$\ln N$}}
\put(264,158){\special{em:moveto}}
\put(985,158){\special{em:lineto}}
\put(985,787){\special{em:lineto}}
\put(264,787){\special{em:lineto}}
\put(264,158){\special{em:lineto}}
\put(624,832){\makebox(0,0){ }}
\put(394,198){\makebox(0,0){$\scriptscriptstyle +$}}
\put(573,316){\makebox(0,0){$\scriptstyle +$}}
\put(642,383){\makebox(0,0){$\scriptstyle +$}}
\put(711,464){\makebox(0,0){$\scriptstyle +$}}
\put(780,557){\makebox(0,0){$\scriptstyle +$}}
\put(849,661){\makebox(0,0){$\scriptstyle +$}}
\put(917,777){\makebox(0,0){$\scriptstyle +$}}
\put(394,195){\special{em:moveto}}
\put(400,197){\special{em:lineto}}
\put(405,199){\special{em:lineto}}
\put(410,202){\special{em:lineto}}
\put(415,204){\special{em:lineto}}
\put(421,207){\special{em:lineto}}
\put(426,210){\special{em:lineto}}
\put(431,213){\special{em:lineto}}
\put(437,216){\special{em:lineto}}
\put(442,218){\special{em:lineto}}
\put(447,221){\special{em:lineto}}
\put(452,225){\special{em:lineto}}
\put(458,228){\special{em:lineto}}
\put(463,231){\special{em:lineto}}
\put(468,234){\special{em:lineto}}
\put(474,238){\special{em:lineto}}
\put(479,241){\special{em:lineto}}
\put(484,245){\special{em:lineto}}
\put(489,248){\special{em:lineto}}
\put(495,252){\special{em:lineto}}
\put(500,256){\special{em:lineto}}
\put(505,259){\special{em:lineto}}
\put(511,263){\special{em:lineto}}
\put(516,267){\special{em:lineto}}
\put(521,271){\special{em:lineto}}
\put(526,275){\special{em:lineto}}
\put(532,280){\special{em:lineto}}
\put(537,284){\special{em:lineto}}
\put(542,288){\special{em:lineto}}
\put(548,293){\special{em:lineto}}
\put(553,297){\special{em:lineto}}
\put(558,301){\special{em:lineto}}
\put(563,306){\special{em:lineto}}
\put(569,311){\special{em:lineto}}
\put(574,315){\special{em:lineto}}
\put(579,320){\special{em:lineto}}
\put(584,325){\special{em:lineto}}
\put(590,330){\special{em:lineto}}
\put(595,335){\special{em:lineto}}
\put(600,340){\special{em:lineto}}
\put(606,345){\special{em:lineto}}
\put(611,351){\special{em:lineto}}
\put(616,356){\special{em:lineto}}
\put(621,361){\special{em:lineto}}
\put(627,367){\special{em:lineto}}
\put(632,372){\special{em:lineto}}
\put(637,378){\special{em:lineto}}
\put(643,384){\special{em:lineto}}
\put(648,389){\special{em:lineto}}
\put(653,395){\special{em:lineto}}
\put(658,401){\special{em:lineto}}
\put(664,407){\special{em:lineto}}
\put(669,413){\special{em:lineto}}
\put(674,419){\special{em:lineto}}
\put(680,425){\special{em:lineto}}
\put(685,432){\special{em:lineto}}
\put(690,438){\special{em:lineto}}
\put(695,444){\special{em:lineto}}
\put(701,451){\special{em:lineto}}
\put(706,457){\special{em:lineto}}
\put(711,464){\special{em:lineto}}
\put(717,471){\special{em:lineto}}
\put(722,477){\special{em:lineto}}
\put(727,484){\special{em:lineto}}
\put(732,491){\special{em:lineto}}
\put(738,498){\special{em:lineto}}
\put(743,505){\special{em:lineto}}
\put(748,512){\special{em:lineto}}
\put(754,519){\special{em:lineto}}
\put(759,527){\special{em:lineto}}
\put(764,534){\special{em:lineto}}
\put(769,541){\special{em:lineto}}
\put(775,549){\special{em:lineto}}
\put(780,556){\special{em:lineto}}
\put(785,564){\special{em:lineto}}
\put(791,572){\special{em:lineto}}
\put(796,579){\special{em:lineto}}
\put(801,587){\special{em:lineto}}
\put(806,595){\special{em:lineto}}
\put(812,603){\special{em:lineto}}
\put(817,611){\special{em:lineto}}
\put(822,619){\special{em:lineto}}
\put(828,627){\special{em:lineto}}
\put(833,636){\special{em:lineto}}
\put(838,644){\special{em:lineto}}
\put(843,652){\special{em:lineto}}
\put(849,661){\special{em:lineto}}
\put(854,669){\special{em:lineto}}
\put(859,678){\special{em:lineto}}
\put(865,687){\special{em:lineto}}
\put(870,695){\special{em:lineto}}
\put(875,704){\special{em:lineto}}
\put(880,713){\special{em:lineto}}
\put(886,722){\special{em:lineto}}
\put(891,731){\special{em:lineto}}
\put(896,740){\special{em:lineto}}
\put(902,749){\special{em:lineto}}
\put(907,759){\special{em:lineto}}
\put(912,768){\special{em:lineto}}
\put(917,777){\special{em:lineto}}
\end{picture}
\vspace{-2.4cm}

\caption{Mean square extent for $E_2=\frac{1}{2}(X-Y)^2$}
\label{fig:meangauss}
\end{figure}

Fig.~\ref{fig:meangauss}  shows the mean square extent for $E_2$
as a function of $N$. A least square fit to
\begin{equation}
     <X^2>_N ~=~ \alpha'' (\ln N)^2 + \beta (\ln N) + \gamma
\end{equation}
yields $\alpha'' =0.080(4)$. The value for $N=8$ can be calculated
analytically and has been added for curiosity.

\begin{figure}[htb]
\setlength{\unitlength}{0.240900pt}
\ifx\plotpoint\undefined\newsavebox{\plotpoint}\fi
\sbox{\plotpoint}{\rule[-0.175pt]{0.350pt}{0.350pt}}%
\special{em:linewidth 0.3pt}%
\begin{picture}(1049,900)(200,-100)
\tenrm
\put(264,158){\special{em:moveto}}
\put(284,158){\special{em:lineto}}
\put(985,158){\special{em:moveto}}
\put(965,158){\special{em:lineto}}
\put(242,158){\makebox(0,0)[r]{3}}
\put(264,263){\special{em:moveto}}
\put(284,263){\special{em:lineto}}
\put(985,263){\special{em:moveto}}
\put(965,263){\special{em:lineto}}
\put(242,263){\makebox(0,0)[r]{4}}
\put(264,368){\special{em:moveto}}
\put(284,368){\special{em:lineto}}
\put(985,368){\special{em:moveto}}
\put(965,368){\special{em:lineto}}
\put(242,368){\makebox(0,0)[r]{5}}
\put(264,473){\special{em:moveto}}
\put(284,473){\special{em:lineto}}
\put(985,473){\special{em:moveto}}
\put(965,473){\special{em:lineto}}
\put(242,473){\makebox(0,0)[r]{6}}
\put(264,577){\special{em:moveto}}
\put(284,577){\special{em:lineto}}
\put(985,577){\special{em:moveto}}
\put(965,577){\special{em:lineto}}
\put(242,577){\makebox(0,0)[r]{7}}
\put(264,682){\special{em:moveto}}
\put(284,682){\special{em:lineto}}
\put(985,682){\special{em:moveto}}
\put(965,682){\special{em:lineto}}
\put(242,682){\makebox(0,0)[r]{8}}
\put(264,787){\special{em:moveto}}
\put(284,787){\special{em:lineto}}
\put(985,787){\special{em:moveto}}
\put(965,787){\special{em:lineto}}
\put(242,787){\makebox(0,0)[r]{9}}
\put(264,158){\special{em:moveto}}
\put(264,178){\special{em:lineto}}
\put(264,787){\special{em:moveto}}
\put(264,767){\special{em:lineto}}
\put(264,113){\makebox(0,0){100}}
\put(373,158){\special{em:moveto}}
\put(373,168){\special{em:lineto}}
\put(373,787){\special{em:moveto}}
\put(373,777){\special{em:lineto}}
\put(436,158){\special{em:moveto}}
\put(436,168){\special{em:lineto}}
\put(436,787){\special{em:moveto}}
\put(436,777){\special{em:lineto}}
\put(481,158){\special{em:moveto}}
\put(481,168){\special{em:lineto}}
\put(481,787){\special{em:moveto}}
\put(481,777){\special{em:lineto}}
\put(516,158){\special{em:moveto}}
\put(516,168){\special{em:lineto}}
\put(516,787){\special{em:moveto}}
\put(516,777){\special{em:lineto}}
\put(545,158){\special{em:moveto}}
\put(545,168){\special{em:lineto}}
\put(545,787){\special{em:moveto}}
\put(545,777){\special{em:lineto}}
\put(569,158){\special{em:moveto}}
\put(569,168){\special{em:lineto}}
\put(569,787){\special{em:moveto}}
\put(569,777){\special{em:lineto}}
\put(590,158){\special{em:moveto}}
\put(590,168){\special{em:lineto}}
\put(590,787){\special{em:moveto}}
\put(590,777){\special{em:lineto}}
\put(608,158){\special{em:moveto}}
\put(608,168){\special{em:lineto}}
\put(608,787){\special{em:moveto}}
\put(608,777){\special{em:lineto}}
\put(625,158){\special{em:moveto}}
\put(625,178){\special{em:lineto}}
\put(625,787){\special{em:moveto}}
\put(625,767){\special{em:lineto}}
\put(625,113){\makebox(0,0){1000}}
\put(733,158){\special{em:moveto}}
\put(733,168){\special{em:lineto}}
\put(733,787){\special{em:moveto}}
\put(733,777){\special{em:lineto}}
\put(797,158){\special{em:moveto}}
\put(797,168){\special{em:lineto}}
\put(797,787){\special{em:moveto}}
\put(797,777){\special{em:lineto}}
\put(842,158){\special{em:moveto}}
\put(842,168){\special{em:lineto}}
\put(842,787){\special{em:moveto}}
\put(842,777){\special{em:lineto}}
\put(876,158){\special{em:moveto}}
\put(876,168){\special{em:lineto}}
\put(876,787){\special{em:moveto}}
\put(876,777){\special{em:lineto}}
\put(905,158){\special{em:moveto}}
\put(905,168){\special{em:lineto}}
\put(905,787){\special{em:moveto}}
\put(905,777){\special{em:lineto}}
\put(929,158){\special{em:moveto}}
\put(929,168){\special{em:lineto}}
\put(929,787){\special{em:moveto}}
\put(929,777){\special{em:lineto}}
\put(950,158){\special{em:moveto}}
\put(950,168){\special{em:lineto}}
\put(950,787){\special{em:moveto}}
\put(950,777){\special{em:lineto}}
\put(969,158){\special{em:moveto}}
\put(969,168){\special{em:lineto}}
\put(969,787){\special{em:moveto}}
\put(969,777){\special{em:lineto}}
\put(985,158){\special{em:moveto}}
\put(985,178){\special{em:lineto}}
\put(985,787){\special{em:moveto}}
\put(985,767){\special{em:lineto}}
\put(985,113){\makebox(0,0){$\ln N$}}
\put(264,158){\special{em:moveto}}
\put(985,158){\special{em:lineto}}
\put(985,787){\special{em:lineto}}
\put(264,787){\special{em:lineto}}
\put(264,158){\special{em:lineto}}
\put(317,189){\raisebox{-1.2pt}{\makebox(0,0){$\Diamond$}}}
\put(489,340){\raisebox{-1.2pt}{\makebox(0,0){$\Diamond$}}}
\put(661,519){\raisebox{-1.2pt}{\makebox(0,0){$\Diamond$}}}
\put(833,719){\raisebox{-1.2pt}{\makebox(0,0){$\Diamond$}}}
\put(317,179){\special{em:moveto}}
\put(317,200){\special{em:lineto}}
\put(307,179){\special{em:moveto}}
\put(327,179){\special{em:lineto}}
\put(307,200){\special{em:moveto}}
\put(327,200){\special{em:lineto}}
\put(489,330){\special{em:moveto}}
\put(489,351){\special{em:lineto}}
\put(479,330){\special{em:moveto}}
\put(499,330){\special{em:lineto}}
\put(479,351){\special{em:moveto}}
\put(499,351){\special{em:lineto}}
\put(661,504){\special{em:moveto}}
\put(661,534){\special{em:lineto}}
\put(651,504){\special{em:moveto}}
\put(671,504){\special{em:lineto}}
\put(651,534){\special{em:moveto}}
\put(671,534){\special{em:lineto}}
\put(833,699){\special{em:moveto}}
\put(833,739){\special{em:lineto}}
\put(823,699){\special{em:moveto}}
\put(843,699){\special{em:lineto}}
\put(823,739){\special{em:moveto}}
\put(843,739){\special{em:lineto}}
\put(317,190){\special{em:moveto}}
\put(322,195){\special{em:lineto}}
\put(327,199){\special{em:lineto}}
\put(332,203){\special{em:lineto}}
\put(338,207){\special{em:lineto}}
\put(343,212){\special{em:lineto}}
\put(348,216){\special{em:lineto}}
\put(353,220){\special{em:lineto}}
\put(358,225){\special{em:lineto}}
\put(364,229){\special{em:lineto}}
\put(369,234){\special{em:lineto}}
\put(374,238){\special{em:lineto}}
\put(379,243){\special{em:lineto}}
\put(384,247){\special{em:lineto}}
\put(390,252){\special{em:lineto}}
\put(395,256){\special{em:lineto}}
\put(400,261){\special{em:lineto}}
\put(405,265){\special{em:lineto}}
\put(410,270){\special{em:lineto}}
\put(416,275){\special{em:lineto}}
\put(421,279){\special{em:lineto}}
\put(426,284){\special{em:lineto}}
\put(431,289){\special{em:lineto}}
\put(437,294){\special{em:lineto}}
\put(442,298){\special{em:lineto}}
\put(447,303){\special{em:lineto}}
\put(452,308){\special{em:lineto}}
\put(457,313){\special{em:lineto}}
\put(463,318){\special{em:lineto}}
\put(468,322){\special{em:lineto}}
\put(473,327){\special{em:lineto}}
\put(478,332){\special{em:lineto}}
\put(483,337){\special{em:lineto}}
\put(489,342){\special{em:lineto}}
\put(494,347){\special{em:lineto}}
\put(499,352){\special{em:lineto}}
\put(504,357){\special{em:lineto}}
\put(510,362){\special{em:lineto}}
\put(515,367){\special{em:lineto}}
\put(520,373){\special{em:lineto}}
\put(525,378){\special{em:lineto}}
\put(530,383){\special{em:lineto}}
\put(536,388){\special{em:lineto}}
\put(541,393){\special{em:lineto}}
\put(546,398){\special{em:lineto}}
\put(551,404){\special{em:lineto}}
\put(556,409){\special{em:lineto}}
\put(562,414){\special{em:lineto}}
\put(567,420){\special{em:lineto}}
\put(572,425){\special{em:lineto}}
\put(577,430){\special{em:lineto}}
\put(583,436){\special{em:lineto}}
\put(588,441){\special{em:lineto}}
\put(593,447){\special{em:lineto}}
\put(598,452){\special{em:lineto}}
\put(603,457){\special{em:lineto}}
\put(609,463){\special{em:lineto}}
\put(614,469){\special{em:lineto}}
\put(619,474){\special{em:lineto}}
\put(624,480){\special{em:lineto}}
\put(629,485){\special{em:lineto}}
\put(635,491){\special{em:lineto}}
\put(640,497){\special{em:lineto}}
\put(645,502){\special{em:lineto}}
\put(650,508){\special{em:lineto}}
\put(655,514){\special{em:lineto}}
\put(661,519){\special{em:lineto}}
\put(666,525){\special{em:lineto}}
\put(671,531){\special{em:lineto}}
\put(676,537){\special{em:lineto}}
\put(682,543){\special{em:lineto}}
\put(687,548){\special{em:lineto}}
\put(692,554){\special{em:lineto}}
\put(697,560){\special{em:lineto}}
\put(702,566){\special{em:lineto}}
\put(708,572){\special{em:lineto}}
\put(713,578){\special{em:lineto}}
\put(718,584){\special{em:lineto}}
\put(723,590){\special{em:lineto}}
\put(728,596){\special{em:lineto}}
\put(734,602){\special{em:lineto}}
\put(739,608){\special{em:lineto}}
\put(744,614){\special{em:lineto}}
\put(749,620){\special{em:lineto}}
\put(755,627){\special{em:lineto}}
\put(760,633){\special{em:lineto}}
\put(765,639){\special{em:lineto}}
\put(770,645){\special{em:lineto}}
\put(775,651){\special{em:lineto}}
\put(781,658){\special{em:lineto}}
\put(786,664){\special{em:lineto}}
\put(791,670){\special{em:lineto}}
\put(796,677){\special{em:lineto}}
\put(801,683){\special{em:lineto}}
\put(807,689){\special{em:lineto}}
\put(812,696){\special{em:lineto}}
\put(817,702){\special{em:lineto}}
\put(822,709){\special{em:lineto}}
\put(827,715){\special{em:lineto}}
\put(833,722){\special{em:lineto}}
\end{picture}
\vspace{-2.4cm}

\caption{Mean square extent for $E_1=|X-Y|$}
\label{fig:meanabsolut}
\end{figure}
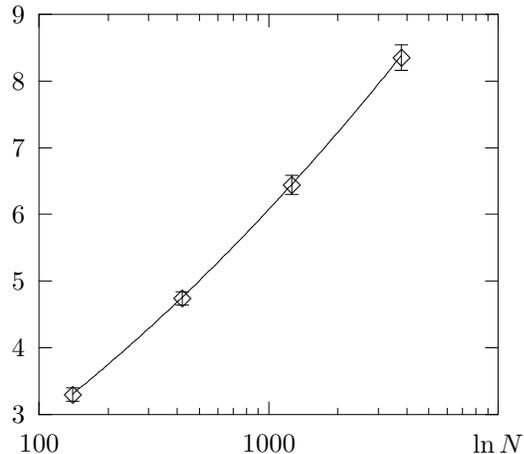

Fig.~\ref{fig:meanabsolut}  shows the
corresponding data for action $E_1$, with a
coefficient $\alpha''=0.11(1)$ from a least square fit. While both actions
quite nicely exhibit the $(\ln N)^2$ behavior, the coefficients are
different, even if we take into account the normalization of the $X$-fields.
This indicates that the coefficient might not be universal.
The disagreement with the matrix model predictions (\ref{matrix})
can be due to the subtraction of graphs with non-trivial 2-point subgraphs.

Figs.~\ref{fig:scale1} and \ref{fig:scale2} show $G(p)$ for the actions
$E_2$ ($N$=34020 triangles) and $E_1$ ($N$=3780 triangles) respectively,
as well as a one-parameter fit with respect to $\langle X^2 \rangle$
using the form (\ref{scale}). The
universality of the scaling function is nicely confirmed. For
smaller lattices the curves, as a function of $p\ln N$, almost are
on top of each other, which again implies the universality of
eq.~\ref{mean}. Small deviations from the theoretical prediction are
not visible in the figures, but can be seen by comparing the numbers.
They are probably due to finite size effects.
{}From the values of $\langle X^2 \rangle$ obtained by fitting $G(p)$,
the coefficient of $(\ln N)^2$
is, consistently with the direct measurement presented before,
1.42 times larger for the action $E_1$ than for $E_2$.
Whether we take into account the flat lattice normalization or not,
this result indicates non-universality of $\alpha''$.

\section{CONCLUSIONS}

The Monte Carlo simulations we performed for the mean square extent
and the tachyonic 2-point function $G(p)$ of $c=1$ matter coupled to
2-dimensonal quantum gravity confirmed the universal scaling behavior
for $G(p)$. Furthermore, they confirmed the
$\langle X^2 \rangle \propto \ln^2(N)$ law and its universality.
The coefficient in front of the $\ln^2$ however
seems not to be universal. The contradiction with matrix model
predictions for this coefficient might be due to a renormalization
of the parameters in the matrix model. The discrepancy between
two $c=1$ scalar fields, which have the same conformal weight in flat
space, can be a more serious problem. Presently, we perform several
numerical checks to clarify these issues.

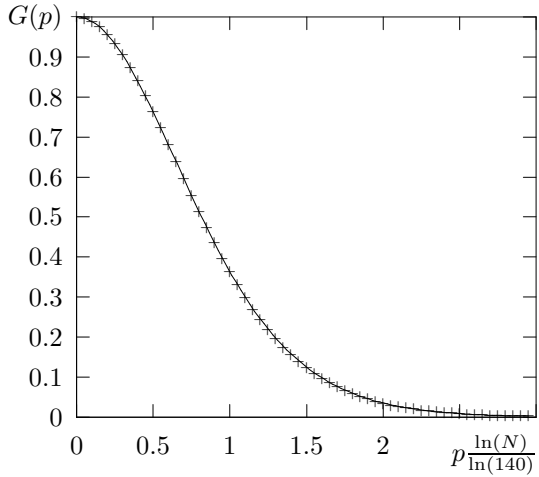
\begin{figure}[htb]
\setlength{\unitlength}{0.240900pt}
\ifx\plotpoint\undefined\newsavebox{\plotpoint}\fi
\sbox{\plotpoint}{\rule[-0.175pt]{0.350pt}{0.350pt}}%
\special{em:linewidth 0.3pt}%
\begin{picture}(1049,900)(150,-100)
\tenrm
\put(264,158){\special{em:moveto}}
\put(985,158){\special{em:lineto}}
\put(264,158){\special{em:moveto}}
\put(264,787){\special{em:lineto}}
\put(264,158){\special{em:moveto}}
\put(284,158){\special{em:lineto}}
\put(985,158){\special{em:moveto}}
\put(965,158){\special{em:lineto}}
\put(242,158){\makebox(0,0)[r]{0}}
\put(264,221){\special{em:moveto}}
\put(284,221){\special{em:lineto}}
\put(985,221){\special{em:moveto}}
\put(965,221){\special{em:lineto}}
\put(242,221){\makebox(0,0)[r]{0.1}}
\put(264,284){\special{em:moveto}}
\put(284,284){\special{em:lineto}}
\put(985,284){\special{em:moveto}}
\put(965,284){\special{em:lineto}}
\put(242,284){\makebox(0,0)[r]{0.2}}
\put(264,347){\special{em:moveto}}
\put(284,347){\special{em:lineto}}
\put(985,347){\special{em:moveto}}
\put(965,347){\special{em:lineto}}
\put(242,347){\makebox(0,0)[r]{0.3}}
\put(264,410){\special{em:moveto}}
\put(284,410){\special{em:lineto}}
\put(985,410){\special{em:moveto}}
\put(965,410){\special{em:lineto}}
\put(242,410){\makebox(0,0)[r]{0.4}}
\put(264,473){\special{em:moveto}}
\put(284,473){\special{em:lineto}}
\put(985,473){\special{em:moveto}}
\put(965,473){\special{em:lineto}}
\put(242,473){\makebox(0,0)[r]{0.5}}
\put(264,535){\special{em:moveto}}
\put(284,535){\special{em:lineto}}
\put(985,535){\special{em:moveto}}
\put(965,535){\special{em:lineto}}
\put(242,535){\makebox(0,0)[r]{0.6}}
\put(264,598){\special{em:moveto}}
\put(284,598){\special{em:lineto}}
\put(985,598){\special{em:moveto}}
\put(965,598){\special{em:lineto}}
\put(242,598){\makebox(0,0)[r]{0.7}}
\put(264,661){\special{em:moveto}}
\put(284,661){\special{em:lineto}}
\put(985,661){\special{em:moveto}}
\put(965,661){\special{em:lineto}}
\put(242,661){\makebox(0,0)[r]{0.8}}
\put(264,724){\special{em:moveto}}
\put(284,724){\special{em:lineto}}
\put(985,724){\special{em:moveto}}
\put(965,724){\special{em:lineto}}
\put(242,724){\makebox(0,0)[r]{0.9}}
\put(264,787){\special{em:moveto}}
\put(284,787){\special{em:lineto}}
\put(985,787){\special{em:moveto}}
\put(965,787){\special{em:lineto}}
\put(242,787){\makebox(0,0)[r]{$G(p)$}}
\put(264,158){\special{em:moveto}}
\put(264,178){\special{em:lineto}}
\put(264,787){\special{em:moveto}}
\put(264,767){\special{em:lineto}}
\put(264,113){\makebox(0,0){0}}
\put(384,158){\special{em:moveto}}
\put(384,178){\special{em:lineto}}
\put(384,787){\special{em:moveto}}
\put(384,767){\special{em:lineto}}
\put(384,113){\makebox(0,0){0.5}}
\put(504,158){\special{em:moveto}}
\put(504,178){\special{em:lineto}}
\put(504,787){\special{em:moveto}}
\put(504,767){\special{em:lineto}}
\put(504,113){\makebox(0,0){1}}
\put(625,158){\special{em:moveto}}
\put(625,178){\special{em:lineto}}
\put(625,787){\special{em:moveto}}
\put(625,767){\special{em:lineto}}
\put(625,113){\makebox(0,0){1.5}}
\put(745,158){\special{em:moveto}}
\put(745,178){\special{em:lineto}}
\put(745,787){\special{em:moveto}}
\put(745,767){\special{em:lineto}}
\put(745,113){\makebox(0,0){2}}
\put(865,158){\special{em:moveto}}
\put(865,178){\special{em:lineto}}
\put(865,787){\special{em:moveto}}
\put(865,767){\special{em:lineto}}
\put(920,100){\makebox(0,0){$p \frac{\ln(N)}{\ln(140)}$}}
\put(985,158){\special{em:moveto}}
\put(985,178){\special{em:lineto}}
\put(985,787){\special{em:moveto}}
\put(985,767){\special{em:lineto}}
\put(264,158){\special{em:moveto}}
\put(985,158){\special{em:lineto}}
\put(985,787){\special{em:lineto}}
\put(264,787){\special{em:lineto}}
\put(264,158){\special{em:lineto}}
\put(264,787){\makebox(0,0){$\scriptscriptstyle +$}}
\put(276,785){\makebox(0,0){$\scriptscriptstyle +$}}
\put(288,780){\makebox(0,0){$\scriptscriptstyle +$}}
\put(300,772){\makebox(0,0){$\scriptscriptstyle +$}}
\put(312,760){\makebox(0,0){$\scriptscriptstyle +$}}
\put(324,745){\makebox(0,0){$\scriptscriptstyle +$}}
\put(336,728){\makebox(0,0){$\scriptscriptstyle +$}}
\put(348,708){\makebox(0,0){$\scriptscriptstyle +$}}
\put(360,687){\makebox(0,0){$\scriptscriptstyle +$}}
\put(372,663){\makebox(0,0){$\scriptscriptstyle +$}}
\put(384,638){\makebox(0,0){$\scriptscriptstyle +$}}
\put(396,613){\makebox(0,0){$\scriptscriptstyle +$}}
\put(408,586){\makebox(0,0){$\scriptscriptstyle +$}}
\put(420,560){\makebox(0,0){$\scriptscriptstyle +$}}
\put(432,533){\makebox(0,0){$\scriptscriptstyle +$}}
\put(444,507){\makebox(0,0){$\scriptscriptstyle +$}}
\put(456,481){\makebox(0,0){$\scriptscriptstyle +$}}
\put(468,456){\makebox(0,0){$\scriptscriptstyle +$}}
\put(480,432){\makebox(0,0){$\scriptscriptstyle +$}}
\put(492,408){\makebox(0,0){$\scriptscriptstyle +$}}
\put(504,387){\makebox(0,0){$\scriptscriptstyle +$}}
\put(516,366){\makebox(0,0){$\scriptscriptstyle +$}}
\put(528,346){\makebox(0,0){$\scriptscriptstyle +$}}
\put(540,328){\makebox(0,0){$\scriptscriptstyle +$}}
\put(552,311){\makebox(0,0){$\scriptscriptstyle +$}}
\put(564,296){\makebox(0,0){$\scriptscriptstyle +$}}
\put(576,282){\makebox(0,0){$\scriptscriptstyle +$}}
\put(588,268){\makebox(0,0){$\scriptscriptstyle +$}}
\put(600,257){\makebox(0,0){$\scriptscriptstyle +$}}
\put(612,246){\makebox(0,0){$\scriptscriptstyle +$}}
\put(625,236){\makebox(0,0){$\scriptscriptstyle +$}}
\put(637,227){\makebox(0,0){$\scriptscriptstyle +$}}
\put(649,219){\makebox(0,0){$\scriptscriptstyle +$}}
\put(661,212){\makebox(0,0){$\scriptscriptstyle +$}}
\put(673,206){\makebox(0,0){$\scriptscriptstyle +$}}
\put(685,200){\makebox(0,0){$\scriptscriptstyle +$}}
\put(697,195){\makebox(0,0){$\scriptscriptstyle +$}}
\put(709,191){\makebox(0,0){$\scriptscriptstyle +$}}
\put(721,187){\makebox(0,0){$\scriptscriptstyle +$}}
\put(733,183){\makebox(0,0){$\scriptscriptstyle +$}}
\put(745,180){\makebox(0,0){$\scriptscriptstyle +$}}
\put(757,177){\makebox(0,0){$\scriptscriptstyle +$}}
\put(769,175){\makebox(0,0){$\scriptscriptstyle +$}}
\put(781,173){\makebox(0,0){$\scriptscriptstyle +$}}
\put(793,171){\makebox(0,0){$\scriptscriptstyle +$}}
\put(805,169){\makebox(0,0){$\scriptscriptstyle +$}}
\put(817,168){\makebox(0,0){$\scriptscriptstyle +$}}
\put(829,167){\makebox(0,0){$\scriptscriptstyle +$}}
\put(841,165){\makebox(0,0){$\scriptscriptstyle +$}}
\put(853,165){\makebox(0,0){$\scriptscriptstyle +$}}
\put(865,164){\makebox(0,0){$\scriptscriptstyle +$}}
\put(877,163){\makebox(0,0){$\scriptscriptstyle +$}}
\put(889,162){\makebox(0,0){$\scriptscriptstyle +$}}
\put(901,162){\makebox(0,0){$\scriptscriptstyle +$}}
\put(913,161){\makebox(0,0){$\scriptscriptstyle +$}}
\put(925,161){\makebox(0,0){$\scriptscriptstyle +$}}
\put(937,161){\makebox(0,0){$\scriptscriptstyle +$}}
\put(949,160){\makebox(0,0){$\scriptscriptstyle +$}}
\put(961,160){\makebox(0,0){$\scriptscriptstyle +$}}
\put(973,160){\makebox(0,0){$\scriptscriptstyle +$}}
\put(271,786){\special{em:moveto}}
\put(278,785){\special{em:lineto}}
\put(285,781){\special{em:lineto}}
\put(293,777){\special{em:lineto}}
\put(300,772){\special{em:lineto}}
\put(307,765){\special{em:lineto}}
\put(314,757){\special{em:lineto}}
\put(321,749){\special{em:lineto}}
\put(328,739){\special{em:lineto}}
\put(336,728){\special{em:lineto}}
\put(343,717){\special{em:lineto}}
\put(350,704){\special{em:lineto}}
\put(357,691){\special{em:lineto}}
\put(364,678){\special{em:lineto}}
\put(371,664){\special{em:lineto}}
\put(379,649){\special{em:lineto}}
\put(386,634){\special{em:lineto}}
\put(393,619){\special{em:lineto}}
\put(400,603){\special{em:lineto}}
\put(407,587){\special{em:lineto}}
\put(414,572){\special{em:lineto}}
\put(422,556){\special{em:lineto}}
\put(429,540){\special{em:lineto}}
\put(436,524){\special{em:lineto}}
\put(443,508){\special{em:lineto}}
\put(450,493){\special{em:lineto}}
\put(457,478){\special{em:lineto}}
\put(465,463){\special{em:lineto}}
\put(472,448){\special{em:lineto}}
\put(479,434){\special{em:lineto}}
\put(486,420){\special{em:lineto}}
\put(493,407){\special{em:lineto}}
\put(500,394){\special{em:lineto}}
\put(507,381){\special{em:lineto}}
\put(515,369){\special{em:lineto}}
\put(522,357){\special{em:lineto}}
\put(529,346){\special{em:lineto}}
\put(536,335){\special{em:lineto}}
\put(543,324){\special{em:lineto}}
\put(550,315){\special{em:lineto}}
\put(558,305){\special{em:lineto}}
\put(565,296){\special{em:lineto}}
\put(572,287){\special{em:lineto}}
\put(579,279){\special{em:lineto}}
\put(586,272){\special{em:lineto}}
\put(593,264){\special{em:lineto}}
\put(601,257){\special{em:lineto}}
\put(608,251){\special{em:lineto}}
\put(615,245){\special{em:lineto}}
\put(622,239){\special{em:lineto}}
\put(629,233){\special{em:lineto}}
\put(636,228){\special{em:lineto}}
\put(644,223){\special{em:lineto}}
\put(651,219){\special{em:lineto}}
\put(658,214){\special{em:lineto}}
\put(665,210){\special{em:lineto}}
\put(672,207){\special{em:lineto}}
\put(679,203){\special{em:lineto}}
\put(687,200){\special{em:lineto}}
\put(694,197){\special{em:lineto}}
\put(701,194){\special{em:lineto}}
\put(708,191){\special{em:lineto}}
\put(715,189){\special{em:lineto}}
\put(722,187){\special{em:lineto}}
\put(729,184){\special{em:lineto}}
\put(737,182){\special{em:lineto}}
\put(744,181){\special{em:lineto}}
\put(751,179){\special{em:lineto}}
\put(758,177){\special{em:lineto}}
\put(765,176){\special{em:lineto}}
\put(772,174){\special{em:lineto}}
\put(780,173){\special{em:lineto}}
\put(787,172){\special{em:lineto}}
\put(794,171){\special{em:lineto}}
\put(801,170){\special{em:lineto}}
\put(808,169){\special{em:lineto}}
\put(815,168){\special{em:lineto}}
\put(823,167){\special{em:lineto}}
\put(830,166){\special{em:lineto}}
\put(837,166){\special{em:lineto}}
\put(844,165){\special{em:lineto}}
\put(851,165){\special{em:lineto}}
\put(858,164){\special{em:lineto}}
\put(866,164){\special{em:lineto}}
\put(873,163){\special{em:lineto}}
\put(880,163){\special{em:lineto}}
\put(887,162){\special{em:lineto}}
\put(894,162){\special{em:lineto}}
\put(901,162){\special{em:lineto}}
\put(909,161){\special{em:lineto}}
\put(916,161){\special{em:lineto}}
\put(923,161){\special{em:lineto}}
\put(930,161){\special{em:lineto}}
\put(937,160){\special{em:lineto}}
\put(944,160){\special{em:lineto}}
\put(951,160){\special{em:lineto}}
\put(959,160){\special{em:lineto}}
\put(966,160){\special{em:lineto}}
\put(973,160){\special{em:lineto}}
\end{picture}
\vspace{-2.4cm}

\caption{$~G(p)$ for $E_2=\frac{1}{2}(X-Y)^2$}
\label{fig:scale1}
\end{figure}
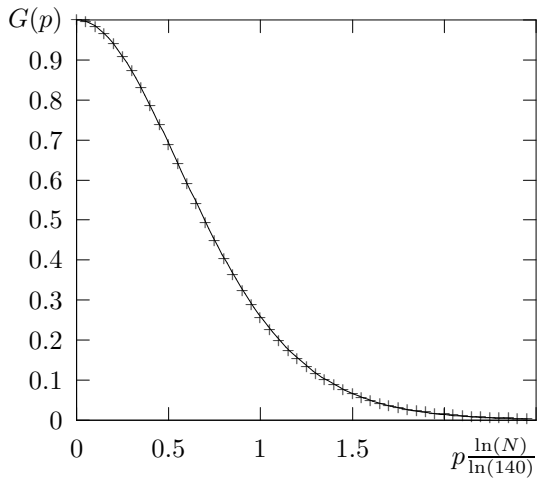
\begin{figure}[htb]
\setlength{\unitlength}{0.240900pt}
\ifx\plotpoint\undefined\newsavebox{\plotpoint}\fi
\begin{picture}(1049,900)(150,-100)
\tenrm
\put(264,158){\special{em:moveto}}
\put(985,158){\special{em:lineto}}
\put(264,158){\special{em:moveto}}
\put(264,787){\special{em:lineto}}
\put(264,158){\special{em:moveto}}
\put(284,158){\special{em:lineto}}
\put(985,158){\special{em:moveto}}
\put(965,158){\special{em:lineto}}
\put(242,158){\makebox(0,0)[r]{0}}
\put(264,221){\special{em:moveto}}
\put(284,221){\special{em:lineto}}
\put(985,221){\special{em:moveto}}
\put(965,221){\special{em:lineto}}
\put(242,221){\makebox(0,0)[r]{0.1}}
\put(264,284){\special{em:moveto}}
\put(284,284){\special{em:lineto}}
\put(985,284){\special{em:moveto}}
\put(965,284){\special{em:lineto}}
\put(242,284){\makebox(0,0)[r]{0.2}}
\put(264,347){\special{em:moveto}}
\put(284,347){\special{em:lineto}}
\put(985,347){\special{em:moveto}}
\put(965,347){\special{em:lineto}}
\put(242,347){\makebox(0,0)[r]{0.3}}
\put(264,410){\special{em:moveto}}
\put(284,410){\special{em:lineto}}
\put(985,410){\special{em:moveto}}
\put(965,410){\special{em:lineto}}
\put(242,410){\makebox(0,0)[r]{0.4}}
\put(264,473){\special{em:moveto}}
\put(284,473){\special{em:lineto}}
\put(985,473){\special{em:moveto}}
\put(965,473){\special{em:lineto}}
\put(242,473){\makebox(0,0)[r]{0.5}}
\put(264,535){\special{em:moveto}}
\put(284,535){\special{em:lineto}}
\put(985,535){\special{em:moveto}}
\put(965,535){\special{em:lineto}}
\put(242,535){\makebox(0,0)[r]{0.6}}
\put(264,598){\special{em:moveto}}
\put(284,598){\special{em:lineto}}
\put(985,598){\special{em:moveto}}
\put(965,598){\special{em:lineto}}
\put(242,598){\makebox(0,0)[r]{0.7}}
\put(264,661){\special{em:moveto}}
\put(284,661){\special{em:lineto}}
\put(985,661){\special{em:moveto}}
\put(965,661){\special{em:lineto}}
\put(242,661){\makebox(0,0)[r]{0.8}}
\put(264,724){\special{em:moveto}}
\put(284,724){\special{em:lineto}}
\put(985,724){\special{em:moveto}}
\put(965,724){\special{em:lineto}}
\put(242,724){\makebox(0,0)[r]{0.9}}
\put(264,787){\special{em:moveto}}
\put(284,787){\special{em:lineto}}
\put(985,787){\special{em:moveto}}
\put(965,787){\special{em:lineto}}
\put(242,787){\makebox(0,0)[r]{$G(p)$}}
\put(264,158){\special{em:moveto}}
\put(264,178){\special{em:lineto}}
\put(264,787){\special{em:moveto}}
\put(264,767){\special{em:lineto}}
\put(264,113){\makebox(0,0){0}}
\put(408,158){\special{em:moveto}}
\put(408,178){\special{em:lineto}}
\put(408,787){\special{em:moveto}}
\put(408,767){\special{em:lineto}}
\put(408,113){\makebox(0,0){0.5}}
\put(552,158){\special{em:moveto}}
\put(552,178){\special{em:lineto}}
\put(552,787){\special{em:moveto}}
\put(552,767){\special{em:lineto}}
\put(552,113){\makebox(0,0){1}}
\put(697,158){\special{em:moveto}}
\put(697,178){\special{em:lineto}}
\put(697,787){\special{em:moveto}}
\put(697,767){\special{em:lineto}}
\put(697,113){\makebox(0,0){1.5}}
\put(841,158){\special{em:moveto}}
\put(841,178){\special{em:lineto}}
\put(841,787){\special{em:moveto}}
\put(841,767){\special{em:lineto}}
\put(920,100){\makebox(0,0){$p \frac{\ln(N)}{\ln(140)}$}}
\put(985,158){\special{em:moveto}}
\put(985,178){\special{em:lineto}}
\put(985,787){\special{em:moveto}}
\put(985,767){\special{em:lineto}}
\put(264,158){\special{em:moveto}}
\put(985,158){\special{em:lineto}}
\put(985,787){\special{em:lineto}}
\put(264,787){\special{em:lineto}}
\put(264,158){\special{em:lineto}}
\put(264,787){\makebox(0,0){$\scriptscriptstyle +$}}
\put(278,785){\makebox(0,0){$\scriptscriptstyle +$}}
\put(293,777){\makebox(0,0){$\scriptscriptstyle +$}}
\put(307,766){\makebox(0,0){$\scriptscriptstyle +$}}
\put(322,750){\makebox(0,0){$\scriptscriptstyle +$}}
\put(336,730){\makebox(0,0){$\scriptscriptstyle +$}}
\put(351,707){\makebox(0,0){$\scriptscriptstyle +$}}
\put(365,681){\makebox(0,0){$\scriptscriptstyle +$}}
\put(379,653){\makebox(0,0){$\scriptscriptstyle +$}}
\put(394,623){\makebox(0,0){$\scriptscriptstyle +$}}
\put(408,592){\makebox(0,0){$\scriptscriptstyle +$}}
\put(423,561){\makebox(0,0){$\scriptscriptstyle +$}}
\put(437,530){\makebox(0,0){$\scriptscriptstyle +$}}
\put(451,499){\makebox(0,0){$\scriptscriptstyle +$}}
\put(466,469){\makebox(0,0){$\scriptscriptstyle +$}}
\put(480,440){\makebox(0,0){$\scriptscriptstyle +$}}
\put(495,412){\makebox(0,0){$\scriptscriptstyle +$}}
\put(509,387){\makebox(0,0){$\scriptscriptstyle +$}}
\put(524,362){\makebox(0,0){$\scriptscriptstyle +$}}
\put(538,340){\makebox(0,0){$\scriptscriptstyle +$}}
\put(552,319){\makebox(0,0){$\scriptscriptstyle +$}}
\put(567,301){\makebox(0,0){$\scriptscriptstyle +$}}
\put(581,284){\makebox(0,0){$\scriptscriptstyle +$}}
\put(596,268){\makebox(0,0){$\scriptscriptstyle +$}}
\put(610,255){\makebox(0,0){$\scriptscriptstyle +$}}
\put(625,242){\makebox(0,0){$\scriptscriptstyle +$}}
\put(639,232){\makebox(0,0){$\scriptscriptstyle +$}}
\put(653,222){\makebox(0,0){$\scriptscriptstyle +$}}
\put(668,214){\makebox(0,0){$\scriptscriptstyle +$}}
\put(682,206){\makebox(0,0){$\scriptscriptstyle +$}}
\put(697,200){\makebox(0,0){$\scriptscriptstyle +$}}
\put(711,194){\makebox(0,0){$\scriptscriptstyle +$}}
\put(725,189){\makebox(0,0){$\scriptscriptstyle +$}}
\put(740,185){\makebox(0,0){$\scriptscriptstyle +$}}
\put(754,181){\makebox(0,0){$\scriptscriptstyle +$}}
\put(769,178){\makebox(0,0){$\scriptscriptstyle +$}}
\put(783,175){\makebox(0,0){$\scriptscriptstyle +$}}
\put(798,173){\makebox(0,0){$\scriptscriptstyle +$}}
\put(812,171){\makebox(0,0){$\scriptscriptstyle +$}}
\put(826,169){\makebox(0,0){$\scriptscriptstyle +$}}
\put(841,168){\makebox(0,0){$\scriptscriptstyle +$}}
\put(855,167){\makebox(0,0){$\scriptscriptstyle +$}}
\put(870,165){\makebox(0,0){$\scriptscriptstyle +$}}
\put(884,164){\makebox(0,0){$\scriptscriptstyle +$}}
\put(898,164){\makebox(0,0){$\scriptscriptstyle +$}}
\put(913,163){\makebox(0,0){$\scriptscriptstyle +$}}
\put(927,162){\makebox(0,0){$\scriptscriptstyle +$}}
\put(942,162){\makebox(0,0){$\scriptscriptstyle +$}}
\put(956,161){\makebox(0,0){$\scriptscriptstyle +$}}
\put(971,161){\makebox(0,0){$\scriptscriptstyle +$}}
\put(271,786){\special{em:moveto}}
\put(278,785){\special{em:lineto}}
\put(285,782){\special{em:lineto}}
\put(293,778){\special{em:lineto}}
\put(300,772){\special{em:lineto}}
\put(307,766){\special{em:lineto}}
\put(314,759){\special{em:lineto}}
\put(321,751){\special{em:lineto}}
\put(328,741){\special{em:lineto}}
\put(335,731){\special{em:lineto}}
\put(343,720){\special{em:lineto}}
\put(350,708){\special{em:lineto}}
\put(357,696){\special{em:lineto}}
\put(364,683){\special{em:lineto}}
\put(371,669){\special{em:lineto}}
\put(378,655){\special{em:lineto}}
\put(385,641){\special{em:lineto}}
\put(392,626){\special{em:lineto}}
\put(400,611){\special{em:lineto}}
\put(407,596){\special{em:lineto}}
\put(414,580){\special{em:lineto}}
\put(421,565){\special{em:lineto}}
\put(428,549){\special{em:lineto}}
\put(435,534){\special{em:lineto}}
\put(442,519){\special{em:lineto}}
\put(450,504){\special{em:lineto}}
\put(457,489){\special{em:lineto}}
\put(464,474){\special{em:lineto}}
\put(471,459){\special{em:lineto}}
\put(478,445){\special{em:lineto}}
\put(485,432){\special{em:lineto}}
\put(492,418){\special{em:lineto}}
\put(500,405){\special{em:lineto}}
\put(507,392){\special{em:lineto}}
\put(514,380){\special{em:lineto}}
\put(521,368){\special{em:lineto}}
\put(528,357){\special{em:lineto}}
\put(535,346){\special{em:lineto}}
\put(542,335){\special{em:lineto}}
\put(549,325){\special{em:lineto}}
\put(557,315){\special{em:lineto}}
\put(564,306){\special{em:lineto}}
\put(571,297){\special{em:lineto}}
\put(578,289){\special{em:lineto}}
\put(585,281){\special{em:lineto}}
\put(592,273){\special{em:lineto}}
\put(599,266){\special{em:lineto}}
\put(607,259){\special{em:lineto}}
\put(614,253){\special{em:lineto}}
\put(621,247){\special{em:lineto}}
\put(628,241){\special{em:lineto}}
\put(635,235){\special{em:lineto}}
\put(642,230){\special{em:lineto}}
\put(649,225){\special{em:lineto}}
\put(657,221){\special{em:lineto}}
\put(664,217){\special{em:lineto}}
\put(671,213){\special{em:lineto}}
\put(678,209){\special{em:lineto}}
\put(685,205){\special{em:lineto}}
\put(692,202){\special{em:lineto}}
\put(699,199){\special{em:lineto}}
\put(707,196){\special{em:lineto}}
\put(714,193){\special{em:lineto}}
\put(721,191){\special{em:lineto}}
\put(728,188){\special{em:lineto}}
\put(735,186){\special{em:lineto}}
\put(742,184){\special{em:lineto}}
\put(749,182){\special{em:lineto}}
\put(756,180){\special{em:lineto}}
\put(764,179){\special{em:lineto}}
\put(771,177){\special{em:lineto}}
\put(778,176){\special{em:lineto}}
\put(785,174){\special{em:lineto}}
\put(792,173){\special{em:lineto}}
\put(799,172){\special{em:lineto}}
\put(806,171){\special{em:lineto}}
\put(814,170){\special{em:lineto}}
\put(821,169){\special{em:lineto}}
\put(828,168){\special{em:lineto}}
\put(835,167){\special{em:lineto}}
\put(842,167){\special{em:lineto}}
\put(849,166){\special{em:lineto}}
\put(856,165){\special{em:lineto}}
\put(864,165){\special{em:lineto}}
\put(871,164){\special{em:lineto}}
\put(878,164){\special{em:lineto}}
\put(885,163){\special{em:lineto}}
\put(892,163){\special{em:lineto}}
\put(899,162){\special{em:lineto}}
\put(906,162){\special{em:lineto}}
\put(913,162){\special{em:lineto}}
\put(921,161){\special{em:lineto}}
\put(928,161){\special{em:lineto}}
\put(935,161){\special{em:lineto}}
\put(942,161){\special{em:lineto}}
\put(949,160){\special{em:lineto}}
\put(956,160){\special{em:lineto}}
\put(963,160){\special{em:lineto}}
\put(971,160){\special{em:lineto}}
\end{picture}
\vspace{-2.4cm}

\caption{~$G(p)$ for $E_1=|X-Y|$}
\label{fig:scale2}
\end{figure}

\section*{ACKNOWLEDGEMENTS}

We would like to express our gratitude
to the HLRZ at KFA J\"ulich where most of our computer runs were performed.
This work was supported in part by the German-Israeli
Foundation for Research and Development (GIF) and
by the Basic Research Foundation of
the Israel Academy of Sciences and Humanities.


\begin{thebibliography}{99}
\bibitem{Polyakov} A. M. Polyakov, Mod.\ Phys.\ Lett.\ A2 (1987) 893.
\bibitem{KPZ}  V. Knizhnik, A. Polyakov and A. Zamolodchikov,
               Mod.\ Phys.\ Lett.\ A3 (1988) 819.
\bibitem{David} F. David, Mod.\ Phys.\ Lett.\ A3 (1988) 1651.
\bibitem{DK}   J. Distler and H. Kawai, Nucl.\ Phys.\ B321 (1989) 509.
\bibitem{Polchinski} J. Polchinski, Nucl.\ Phys.\ B346 (1990) 253.
%
\bibitem{GM} D.~Gross and A.~A.~Migdal, Phys.\ Rev.\ Lett.\ 64 (1990) 127.
\bibitem{DS} M.~R.~Douglas and S.~H.~Shenker, Nucl.\ Phys.\ B335 (1990) 635.
\bibitem{BK} E.~Br\'ezin and V.~A.~Kazakov. Phys.\ Lett.\ B236 (1990) 144.
\bibitem{Gross} D.\ Gross and N.\ Miljkovi\'c, Phys.\ Lett.\ 238B (1990) 217.
\bibitem{Kostov} I. Kostov, Phys.\ Lett.\ 215B (1988) 499.
\bibitem{GKN}    D. J. Gross, I. R. Klebanov and M. J. Newman,
                 Nucl.\ Phys.\ B350 (1991) 621.
%
\bibitem{km} V. Kazakov and A. A. Migdal, Nucl.\ Phys.\ B311 (1989) 171.
\bibitem{Distler} J. Distler, Z. Hlousek and H. Kawai,
                  Int.\ J. Mod.\ Phys.\ A5 (1989) 1093.
\bibitem{VMR}
H. G. Evertz, M. Hasenbusch, M. Marcu, K. Pinn and S. Solomon,\\
Phys.\ Lett.\ 254B (1991) 185,\\
in {\em Lattice 90\/}, Nucl.\ Phys.\ B (Proc.\ Suppl.) 20 (1991) 80, and \\
Int.\ J.\ Mod.\ Phys.\ C3 (1992) 235.
%
\bibitem{KKM} V. A. Kazakov, I. K. Kostov and A. A. Migdal,
                     Phys. Lett. 157B (1985) 295.
\bibitem{BKKM} D. V. Boulatov, V. A. Kazakov, I.~K.~Kostov and A.~A.~Migdal,
               Nucl.\ Phys.\ B275 (1986) 641.
\bibitem{David2}  F. David, Nucl.\ Phys.\ B257 (1985) 45.
\bibitem{Billoire} A. Billoire and F. David, Nucl.\ Phys.\ B275 (1986) 617.
\bibitem{ADF}      J. Ambj\o rn, B. Durhuus and J.~Fr\"ohlich,
                   Nucl.\ Phys.\ B257 (1985) 433.
\end{thebibliography}
\end{document}